%% file: CHASE.tex
\documentclass[10pt,conference]{IEEEtran} 
\IEEEoverridecommandlockouts
\usepackage{cite}
\usepackage{amsmath,amssymb,amsfonts}
\usepackage{algorithmic}
\usepackage{graphicx}
\usepackage{textcomp}
\usepackage{xcolor}
\def\BibTeX{{\rm B\kern-.05em{\sc i\kern-.025em b}\kern-.08em
    T\kern-.1667em\lower.7ex\hbox{E}\kern-.125emX}}
\usepackage{graphicx}
\usepackage{cite}   
\usepackage{xcolor}
\usepackage [english]{babel}
\usepackage [autostyle, english = american]{csquotes}
\MakeOuterQuote{"}
\usepackage{verbatim}
\usepackage{adjustbox}
\usepackage{xcolor}
\usepackage{colortbl}
\usepackage{booktabs}
\usepackage{longtable}
\usepackage{subfig}
\usepackage{tabularx}
\usepackage{multirow}
\usepackage{xltabular}
\usepackage{soul}
\usepackage{float}
\usepackage{placeins}
\usepackage{graphicx}
\usepackage{wrapfig}
\usepackage{lscape}
\usepackage{rotating}
\usepackage{csquotes}
\usepackage{epstopdf}
\usepackage[T1]{fontenc}
\usepackage{upquote}
\usepackage{subfig}
\usepackage{graphicx}
\usepackage{caption}
\usepackage{graphicx}
\usepackage{adjustbox}
\graphicspath{{Charts_figs/}}
\usepackage{adjustbox}
\usepackage{supertabular}     

\makeatletter
\newcommand{\newlineauthors}{%
  \end{@IEEEauthorhalign}\hfill\mbox{}\par
  \mbox{}\hfill\begin{@IEEEauthorhalign}
}

\begin{document}

\title{Emotions in Requirements Engineering: A Systematic Mapping Study\\
}

\author{

\IEEEauthorblockN{Tahira Iqbal}
\IEEEauthorblockA{\textit{Institute of Computer Science, } \\
\textit{University of Tartu,  Estonia} \\
tahira.iqbal@ut.ee}
\and
\IEEEauthorblockN{Hina Anwar}
\IEEEauthorblockA{\textit{Institute of Computer Science, } \\
\textit{University of Tartu,  Estonia} \\
hina.anwar@ut.ee}
\and
\IEEEauthorblockN{Syazwanie Filzah}
\IEEEauthorblockA{\textit{Institute of Computer Science, } \\
\textit{University of Tartu,  Estonia} \\
syazwanie.filzah.binti.zulkifli@ut.ee}

\newlineauthors
\IEEEauthorblockN{Mohamad Gharib}
\IEEEauthorblockA{\textit{Institute of Computer Science, } \\
\textit{University of Tartu,  Estonia} \\
mohamad.gharib@ut.ee}
\and

\IEEEauthorblockN{Kerli Mooses}
\IEEEauthorblockA{\textit{Institute of Computer Science, } \\
\textit{University of Tartu,  Estonia} \\
kerli.mooses@ut.ee}
\and
\IEEEauthorblockN{Kuldar Taveter}
\IEEEauthorblockA{\textit{Institute of Computer Science, } \\
\textit{University of Tartu,  Estonia} \\
kuldar.taveter@ut.ee}
 \thanks{The research work presented in this paper has received funding from the Pilots for Healthy and Active Ageing (Pharaon) project of the European Union’s Horizon 2020 research and innovation programme under the grant agreement no. 857188 and from the European Social Fund via the IT Academy programme.}

}

\maketitle
\begin{abstract}
The purpose of requirements engineering (RE) is to make sure that the expectations and needs of the stakeholders of a software system are met.   Emotional needs can be captured as emotional requirements that represent how the end user should feel when using the system. Differently from functional and quality (non-functional) requirements, emotional requirements have received relatively less attention from the RE community. This study is motivated by the need to explore and map the literature on emotional requirements. The study applies the systematic mapping study technique for surveying and analyzing the available literature to identify the most relevant publications on emotional requirements. We identified 34 publications that address a wide spectrum of practices concerned with engineering emotional requirements. The identified publications were analyzed with respect to the application domains, instruments used for eliciting and artefacts used for representing emotional requirements, and the state of the practice in emotion-related requirements engineering. This analysis serves to identify research gaps and research directions in engineering emotional requirements. To the best of the knowledge by the authors, no other similar study has been conducted on emotional requirements.
\end{abstract}


\newcommand{\tablepath}{Tables}
\input{Introduction.tex}

\input{Study_Design}

\input{Results}

\input{Disscussion.tex}

\input{ThreatsToValidity.tex}
\input{Conclusion.tex}

\bibliographystyle{IEEEtran}
\bibliography{./req}   
\newpage

\appendix
\input{Tables/ref_table}
\end{document}

%% file: Introduction.tex
\section{Introduction}

A key success factor in designing any software system is meeting the expectations by its stakeholders, which is also the purpose of Requirements Engineering (RE) that focuses on eliciting, categorizing,  prioritizing, and validating stakeholders' actual needs that drive the requirements for the system-to-be \cite{Softwareengineering8}. It is well acknowledged in the RE community that requirements can be broadly categorized into functional and non-functional (quality) requirements, and a successful system should satisfy both of these requirements \cite{chung2009non,chung2012non}. 

Although many frameworks and approaches have been developed to deal with various kinds of quality requirements (e.g., security, privacy, reliability), emotional requirements have received relatively less attention from the RE community \cite{curumsing2017emotion, Taveter}. Emotional requirements represent how the user should feel when interacting with the software system \cite{P19}. For example, a patient wants to feel capable of managing her or his own health using an e-healthcare system \cite{P30}. The research results presented in the studies \cite{mendoza2013role,ramos2005emotion,pedellemotions,miller2015emotion,iqbal2023theory} reveal that software engineers fail to give fair consideration to the emotional needs of users when designing systems and point out that emotional needs have not been successfully addressed in the software engineering field. More specifically, emotional requirements by end users are either not considered at all or are vaguely dealt with in requirements engineering \cite{P4}. This may result in a system that does not offer essential features that are required by its user, which, in turn, will hinder the system adoption by the user for whom it has been developed \cite{P8, GeneratingRequirementsOutofThinAir} 

To this end, identifying the emotional requirements of end users and considering them during system design is essential for a successful software system as it improves the system acceptance \cite{mendoza2013role} and reduces the probability of system failure \cite{lopez2014modelling}. Moreover, not giving adequate consideration to such requirements can lead to unhappy and frustrated users \cite{curumsing2019emotion}, which may lead to huge costs because of the need to develop a new system as a result of user rejection \cite{P8}. However, such a problem could be avoided if the emotional requirements would be explicitly captured and addressed during RE and system design \cite{miller2015emotion}.

Currently, there is a lack of software engineering methodologies and frameworks that can deal with emotional requirements \cite{curumsing2017emotion}. Nonetheless, exciting works related to user emotions and emotional requirements are scattered in the literature. In this context, this paper applies the mapping study technique for surveying the available literature to identify the most relevant publications on emotional requirements. This is followed by analyzing the selected publications to answer the research questions.

The rest of this publication is organized as follows. In Section \ref{sec:study_design}, we present the methodology, which includes the research objective and questions, search strategy, selection criteria, quality assessment, data extraction strategy, and process used in this study. Section \ref{sec:results} presents the results of the study. Section \ref{sec:Discussion} discusses the results and threats to the validity of our study. Finally, in Section \ref{sec:conclusion}, the study is concluded, and ideas for future work are presented.

%% file: Study_Design.tex
\section{Study Design}
\label{sec:study_design}

\subsection {Research questions}
\label{sec:research_questions}
Our RQs have been designed with the objective to examine the current research literature on emotions in RE, starting from providing an overview of the state-of-the-art literature and proceedings with focusing on the current practices. The research questions (RQs) for our study are as follows:

\begin{itemize}
    \item{ \textbf{RQ1.} What is the current state of the literature on emotions in RE?}
    \item{ \textbf{RQ2.} How are emotions elicited and represented in different phases of RE?}
    \item{ \textbf{RQ3.} What is the state of the practice in emotion-related RE?}
\end{itemize}

\subsection{Research method}
For our research method, we followed the guidelines proposed by Petersen et al. \cite{petersen2008systematic} and Kitchenham et al. \cite{kitchenham2007cross}. Our combined research method includes three phases: publication search, publication selection, and data extraction. These phases are described in the following numbered subsections:

\subsubsection{Publication search} 

In the first phase of the mapping study, i.e., publication search, we followed three steps to collect publications for our mapping study. These steps consisted of database search, query formulation, and duplicate removal, explained in the following paragraphs:

\noindent\textbf{\textit{Database search:}} We selected the following digital libraries as the data sources for our study: ACM Digital Library (ACM)\footnote{https://dl.acm.org/search/advanced}, IEEE Xplore Digital Library (Xplore)\footnote{https://ieeexplore.ieee.org/search/advanced}, Web of Science (WOS)\footnote{https://webofknowledge.com}, Scopus\footnote{https://www.scopus.com/search/form.uri?display=advanced}, and SpringerLink (SL)\footnote{https://link.springer.com/advanced-search}. We chose these libraries as per the recommendations in the guidelines by Kitchenham, et al. \cite{kitchenham2007guidelines} and Petersen, et al. \cite{petersen2015guidelines}. We expanded our data sources to the additional libraries by ACM and Springer because they host major journals and conference proceedings on RE and software engineering \cite{zhao2021natural}.

\noindent\textbf{\textit{Query formulation:}}
We formulated the search query to search the aforementioned digital libraries. The search terms for querying these libraries were constructed in two steps. In the first step, we followed the approach presented in the literature by Kitchenham, et al. \cite{kitchenham2007cross}. 

In the second step, we analyzed the existing mapping studies performed in the field of RE and emotions. After that, we extracted the keywords used by those existing studies. Next, we complemented our search string with those keywords that were relevant and not already included in our search string. In our case, the search query contained the base terms `requirements engineering' and `emotion'. For the first base term, i.e., `requirements engineering', we searched the existing mapping studies and systematic literature reviews (SLR) with this base term and then extracted the keywords that were used in the search strings of those studies together with the base term. After that, we examined the resulting keywords and added them to our search string if they were relevant and not already included in our search string. We repeated the same process for elaborating the `emotion' base term. Following, we linked all of the retrieved keywords by the use of boolean expressions to obtain more search results. First, we combined the keywords of both base terms with the `OR' logical operator. Later, we connected the keywords of the base terms by means of the logical `AND' operator. We optimized and refined our preliminary search string during multiple iterations and discussions because it is essential to formulate a comprehensive query for performing an intensive and widespread search. The complete search query is presented in Table \ref{query_table}.

\input{Tables/query_table}

\noindent\textbf{\textit{Duplicates' removal:}}
After searching the libraries with the query, we downloaded the results and presented the results for each library as a single Google sheet. To handle duplicates, we used the "Data Cleanup" feature of Google sheets, which identifies possible duplicate publications based on a selection of columns. In our case, the relevant columns were `title', `authors', `year', and `DOI', and in case the values in all of those columns were identical for a publication, the duplicate was identified, and the publication was removed from the corresponding sheet. To remove duplicates among all of the libraries, we merged the Google sheets of all of the libraries and applied the "Data Cleanup" feature again. The number of publications for each library after removing duplicates is shown in Table \ref{tab:my-table}.\footnote{https://tinyurl.com/56z4y4ud}

\begin{table}[H]
\caption{Numbers of publications after removing duplicates}
\label{tab:my-table}
\centering
\begin{tabularx}{.49\textwidth}{X | c | c| c| c| c | c}
\hline
\textbf{Library} &Springer& ACM  & Scopus  & WOS & IEEE &  \textbf{Total} \\ \hline
\textbf{No. of publications}    &  2935 & 600  &  221     & 105 & 80 & \textbf{3941} \\ \hline
\end{tabularx}

\end{table}

\subsubsection{Publication selection}
The publication selection phase consisted of four steps: applying inclusion and exclusion criteria, title abstract selection, abstract selection and quality assessment, discussed in the following paragraphs.

\noindent\textbf{\textit{Inclusion and exclusion criteria:}} For performing step (i), we first defined a set of inclusion and exclusion criteria for our publication selection. These criteria are shown in Table \ref{tab:inclusion_exclusion}. According to the criteria, we included peer-reviewed articles published between 2000 and 2021. We started from the publishing year 2000 because our query yielded only a few results before this year; we also wanted to keep the focus on recent years due to the increased adoption of advanced technologies by the industry.

\input{Tables/Inclusion_exclusion}

\noindent\textbf{\textit{Title selection:}}
For performing step (ii), we organized the participating researchers into three groups, where each group consisted of two researchers. Each group member read the title and then independently labelled the publication as "include", "exclude", or "unsure". Later, the labelled results were matched, and conflicts were resolved between the groups after discussions. At this point, abstracts were also read for a better understanding of the articles. 
The outcome of the publication selection phase after performing both steps (i) and (ii) was 359 publications.

\input{Tables/Quality_criteria}
\noindent\textbf{\textit{Abstract selection:}}
For performing step (iii), we followed the same process performed in Title selection (ii). However,  each group member read the abstract instead of the title in this step to label the publication as "include", "exclude", or "unsure", and conflicts were resolved after discussion. The outcome of the abstract selection step was 96 publications.

\noindent\textbf{\textit{Quality assessment}}: Using the inclusion and exclusion criteria as well as reading titles and abstracts of the study can help with reducing the noise by excluding non-relevant publications, but it neither guarantees that the selected studies are of high quality nor that they are relevant to the scope of our research questions. With this consideration, we applied quality assessment (QA) criteria to the publications that resulted from the publication selection phase (96 publications in total) to identify the most relevant, complete, and high-quality publications for answering our research questions. In particular, we formulated the five quality assessment questions shown in Table \ref{tab:quality_criteria} to identify: 1- the most relevant studies to answer our research questions (Q1); 2- whether such studies have high-quality content that we can use to answer our research questions (Q2); 3- whether such studies have been applied in a realistic setting or they are purely theoretical (Q3); 4- whether such studies can be considered complete, i.e., their results have been validated (Q4); and whether their limitations and/or threats to validity have been discussed (Q5). 
\begin{figure}[!t]
\centering
\includegraphics[width=0.7\linewidth]{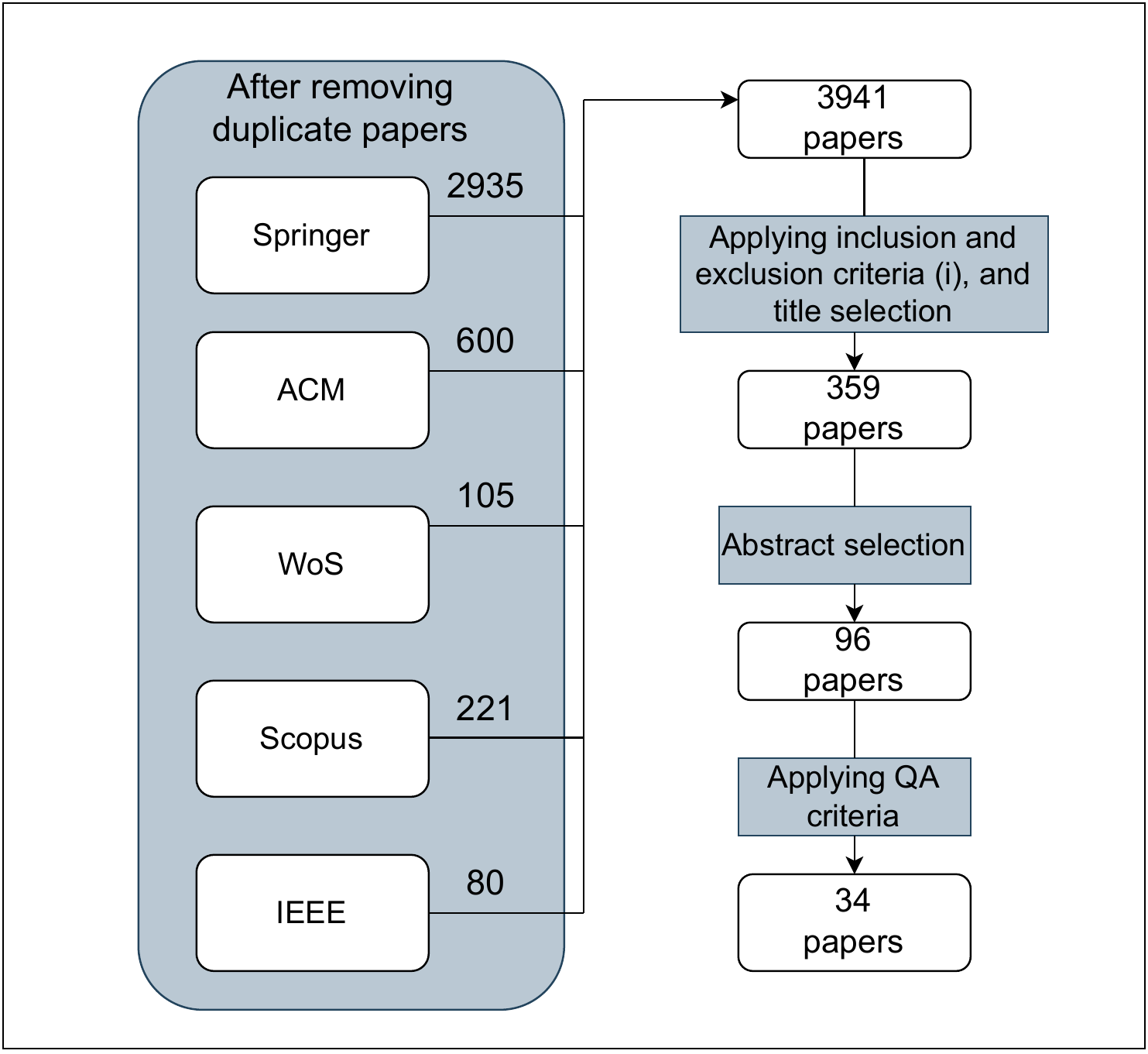}
\caption{Search and selection process of primary articles}
\label{fig:PrimaryStudiesSelection}
\end{figure}

\input{Tables/Quality_Scores}

The application of QA was performed in two groups, each consisting of two researchers. We divided the publications into two parts of equal size and assigned one part to each group. After examining the publication, each group member independently assigned a score between 0 and 1 to each question. Then, the quality score (QS) was computed for the publication by summing the scores of all QA questions. Later, we calculated the average QS for each group, and the publication was only included if the average score was at least 3. The decision to choose 3 as a cutoff point for a study selection was made by the authors after performing a preliminary analysis to identify the most appropriate cutoff point that will guarantee excluding any study that is irrelevant or has limited relevance to our study. More specifically, the authors have randomly selected 18 studies (3 for each author), applied the QA criteria, and calculated the QS for each of these studies. After that, the authors discussed the relationship between the assigned score and the content of each of these studies, which led to the unanimous selection of 3 as a cutoff point. Table \ref{tab:QualityScore} shows the quality scores that have been assigned to publications. This was the final reduction activity, which resulted in 34 primary publications, the references to which are presented in Table \ref{tab:ref_table} in the Appendix. Figure ~\ref{fig:PrimaryStudiesSelection} depicts the process of selecting primary publications.

\subsubsection{Data extraction}
\label{sec:Extraction}
Following the publication selection process, we extracted the information about the final set of publications into a Google sheet. In case any disagreements occurred during the data extraction, they were discussed and resolved among the authors.

To answer RQ1, we extracted information about the publication year, venue, and content type to classify publications based on the publication year and publication type. We further identified domain categories of publications by using a bottom-up merging technique. According to this technique, we read each publication's abstract, introduction, method, and results' sections to extract the keywords for identifying the domain category of the publication. After discussion among the researchers, domain categories were decided based on the keywords. The resulting domain categories were used to classify the publications. 
 
To answer RQ2, we read the introduction, research method, and results' sections of the selected publications to identify the relevant RE phases and associated RE problems explored in the publications. To avoid conflicts, we first conducted a trial session with all the involved researchers to create a mutual understanding of how to identify instruments used for eliciting and representing emotional requirements. After that, each researcher extracted keywords to identify the instruments used in the process of engineering emotional requirements. After discussion among the authors, the instrument categories were decided based on the keywords. The identified instrument categories were used to classify the publications. 
 
To answer RQ3, we read the research method and results' sections of the selected publications to identify new tools (if any) created to help practitioners in the process of engineering emotional requirements. We also collected data regarding the availability of a tool, i.e., if the tool is open source or not. Additionally, we extracted information about various artefacts, such as models, methods and documents, that are generated or modified according to the given publication to facilitate the process of engineering emotional requirements. Similarly to RQ1 and RQ2, we adopted a bottom-up merging technique to form categories of the identified artefacts.

%% file: Tables/query_table.tex
\begin{table}[!htbp]
\centering
\caption{Search query}
\label{query_table}
\begin{tabularx}{.48\textwidth}{|p{8.3cm}|}

\hline
("requirements engineering" OR "requirements elicitation"  OR "requirements analysis"  OR "requirements specification" OR "requirements modelling" OR "requirements validation"  OR "requirements verification" OR "requirements management"OR "emotional requirements"  OR "affective requirements" OR "value-based requirements engineering")\\
\hline
\multicolumn{1}{|c|}{AND} \\ \hline
("emotions" OR "sentiment" OR "emotional" OR "sentiment analysis" OR "semiotics"  OR "emoji" OR "emotion recognition" OR "affective feedback" OR "affective state detection" OR "cognitive state detection" OR "emotion classification" OR "human factor"  OR "human values" OR "motivational modelling")\\
\hline
\end{tabularx}
\end{table}

%% file: Tables/Inclusion_exclusion.tex
\begin{table}[!t]
\caption{(I)nclusion and (E)xclusion criteria for selecting relevant publications}
\label{tab:inclusion_exclusion}
\centering
\captionsetup{justification=centering}
\begin{tabularx}{.48\textwidth}{c|c|X}
\hline
\textbf{ I/E } & \textbf{ No. } & \multicolumn{1}{c}{\textbf{Criteria}} \\ \hline
I & 1 & Include peer-reviewed primary publications \\ \hline
I & 2 & Include articles published between 2000 and 2021 \\ \hline
E & 1 & Exclude tables of contents, editorials, white papers, standards, extended abstracts, books, tutorials, standards, non-peer-reviewed publications, and duplicate articles.\\ \hline
E & 2 & Articles that are not published in English \\ \hline
\end{tabularx}

\end{table}

%% file: Tables/Quality_criteria.tex
\begin{table}[!b]
\centering
\caption{Quality assessment criteria}
\label{tab:quality_criteria}
\begin{tabularx}{.48\textwidth}{c|X|c}

    \toprule
   
   \textbf{ ID} &	\multicolumn{1}{|c|}{\textbf{Quality assessment question}} & \textbf{Score}\\
    \midrule
\textbf{Q1} &   Is the content/contribution of the work related to RQs of our mapping study? & 1\\ \hline
\textbf{Q2} &  Are the objectives/needs of the work clearly justified and aligned with the RQs of our study? & 1\\ \hline 
\textbf{Q3} &   Has the work (solution/artefact) been applied in the industry or justified by an appropriate case study/project? & 1\\ \hline
\textbf{Q4} &  Does the work contain validation of results? & 1 \\ \hline
\textbf{Q5}  &  Does the work discuss limitations and/or threats to validity? & 1\\  \hline
&	\multicolumn{1}{|c|}{\textbf{TOTAL}}        &	\textbf{5}\\
\bottomrule
\end{tabularx}

\end{table}

%% file: Tables/Quality_Scores.tex
\begin{table}[!b]
    \centering
   \captionof{table}{Quality scores assigned to publications}
       \label{tab:QualityScore}
    \begin{tabularx}{.34\textwidth}{X| c} 
    \hline
  
   Publication ID	& Quality Score \\\hline
   P23, P29, P31, P32 &	3\\\hline
   P16, P27 &	3.25\\\hline
   P5, P7, P11, P18, P21, P34 &	3.5\\\hline
   P13, P26, P28, P33 &	3.75\\\hline
   P1, P9, P10, P14, P20, P30 &	4\\\hline
   P2, P15, P17, P22, P25 &	4.25\\\hline
    P3, P6, P8  &	4.5\\\hline
  P12, P19  &	4.75\\\hline
   P4, P24 &	5\\\hline
    \end{tabularx}

\end{table}

%% file: Results.tex
\section{Results}
\label{sec:results}
In this section, we present the results of our study. The list of the selected publications and additional details are shown in Annex A. Subsections ~\ref{sec:4.1}, ~\ref{sec:4.2}, and ~\ref{sec:4.3} are dedicated to answering the research questions. 

\subsection{\textbf{RQ1.} What is the state of the literature on emotions in RE?}
\label{sec:4.1}

\begin{figure}[!hbt]
\centering
\includegraphics[width=\columnwidth]{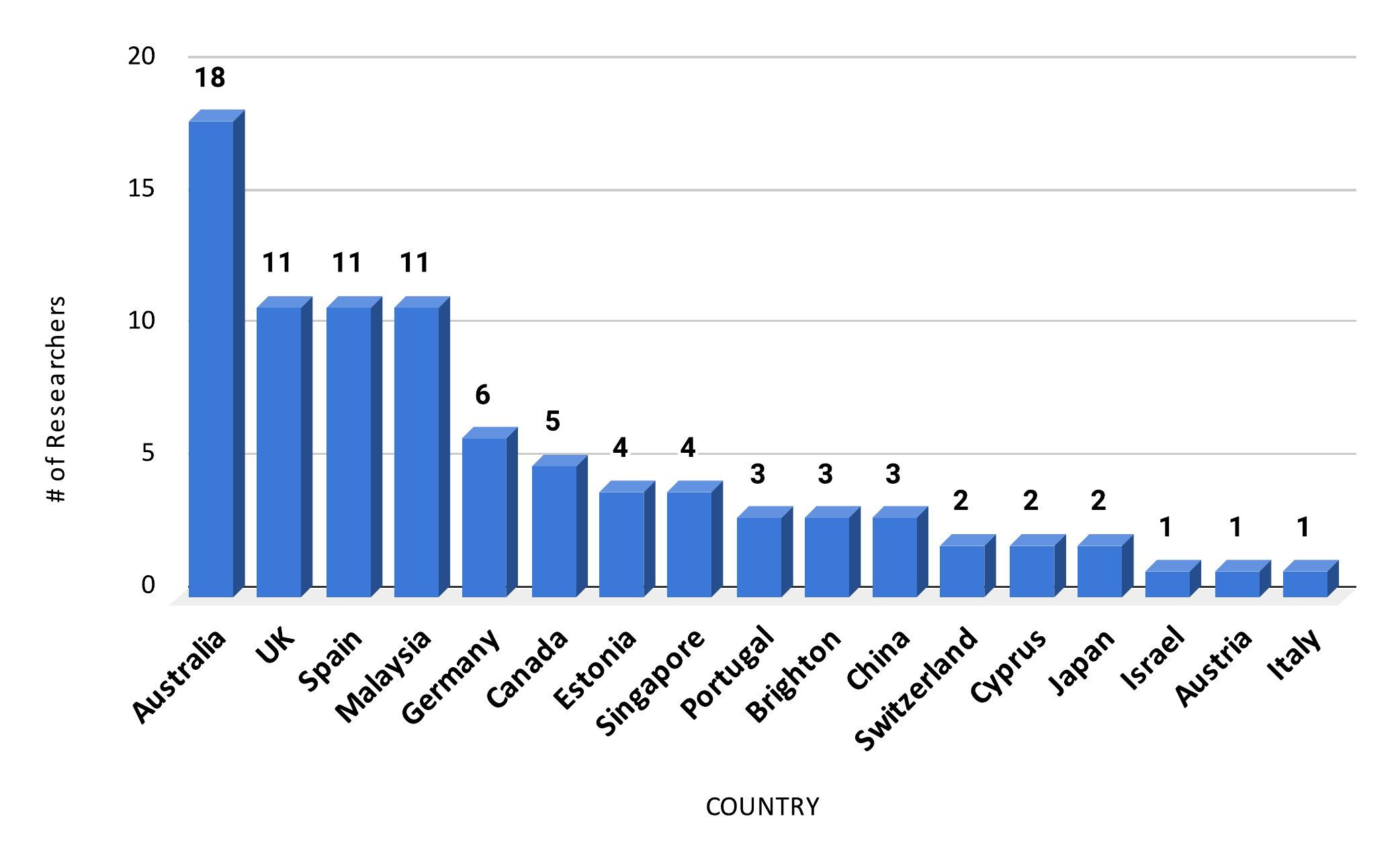}
\caption{Distribution of researchers per geographical location}
\label{fig:authorlocations}
\end{figure}
\begin{figure*}[!t]
  \centering
  \begin{minipage}[b]{0.5\textwidth}
    \includegraphics[width=\textwidth]{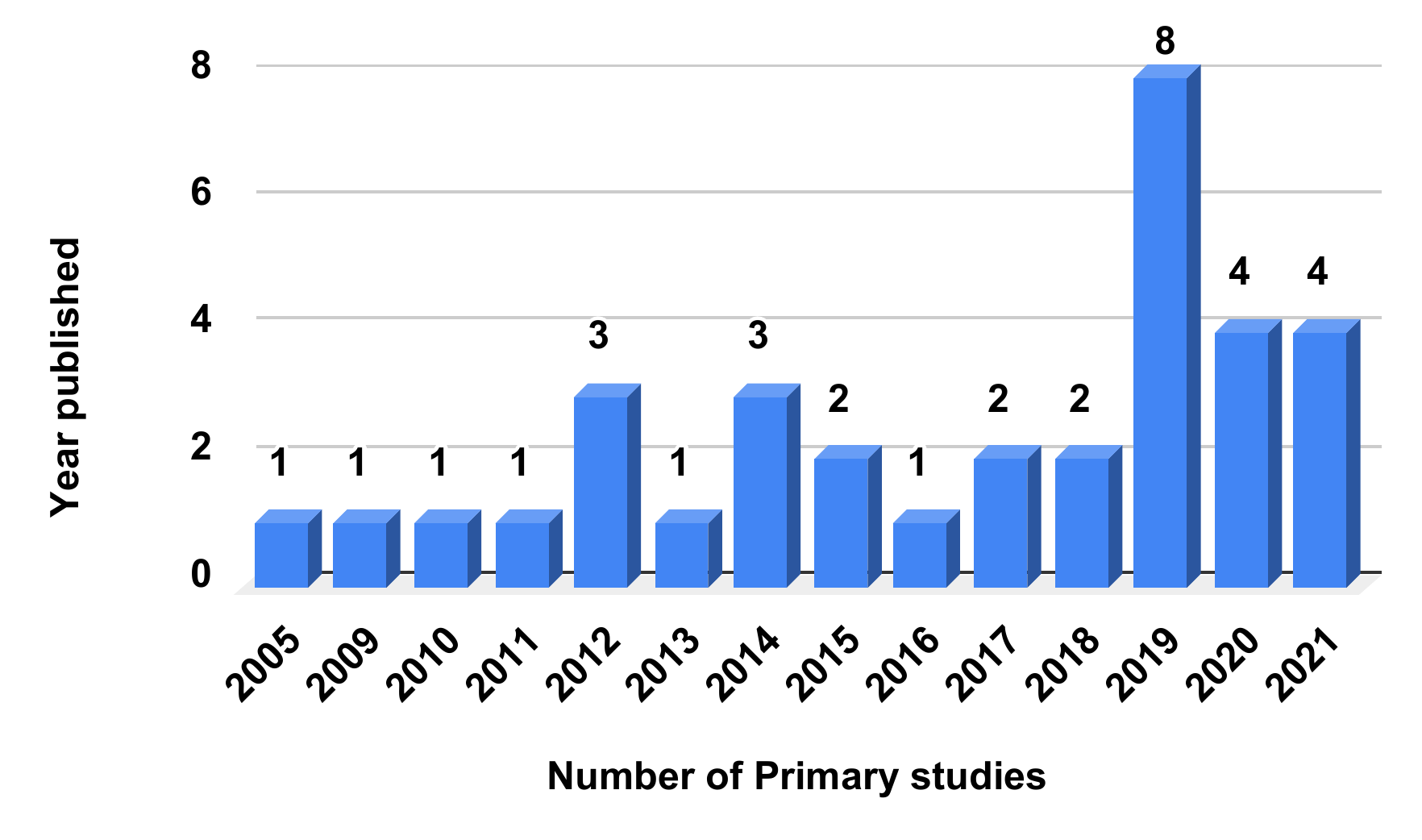}
    \caption{Number of primary publications published by year}
    \label{fig:2figsA}
  \end{minipage}
  \hfill
  \begin{minipage}[b]{0.49\textwidth}
    \includegraphics[width=\textwidth]{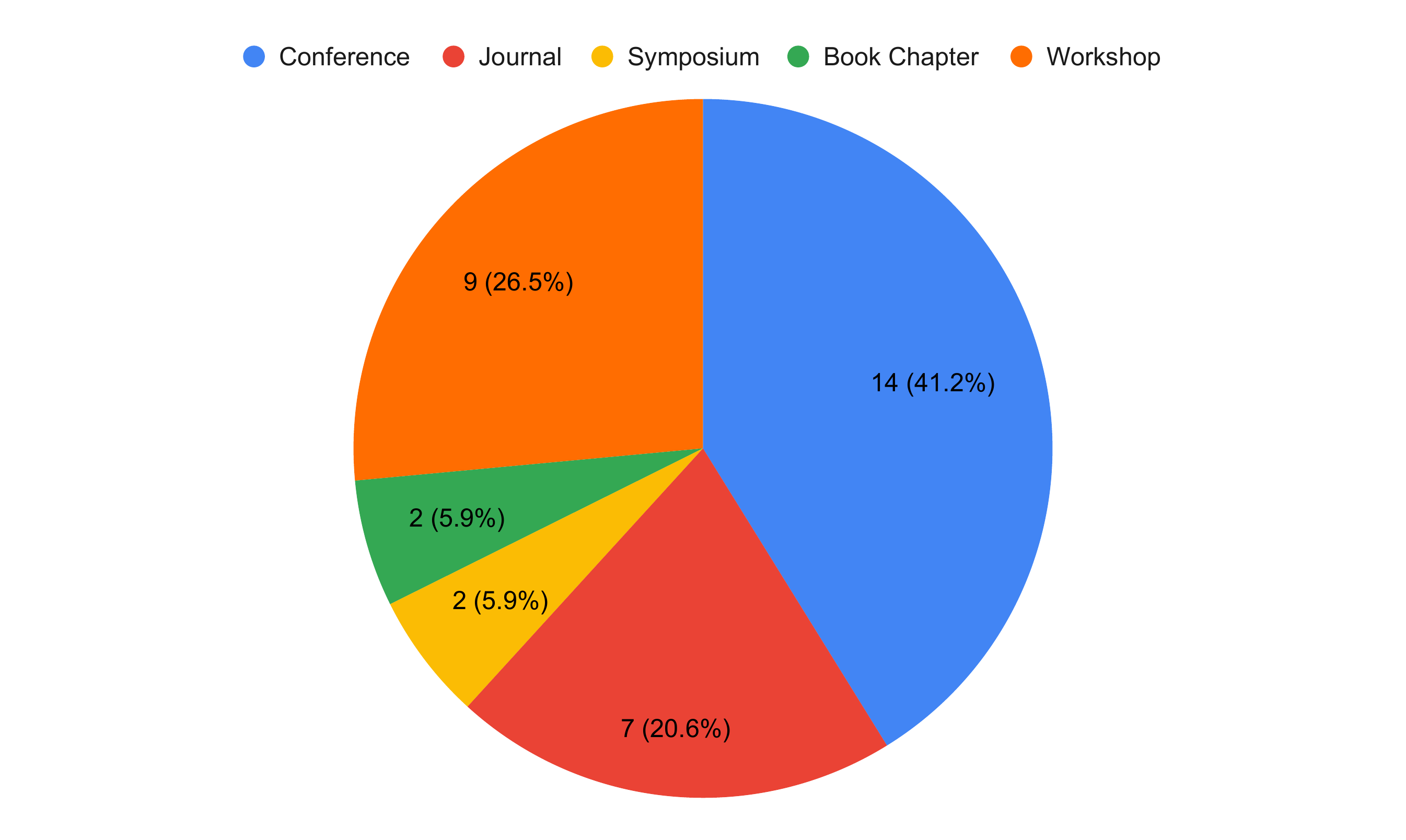}
    \caption{Proportions of articles by publication type}
		\label{fig:2figsB}
  \end{minipage}
\end{figure*}

 
We categorized the selected publications with regard to the publication type, year of publication, type of venue, geographical location, affiliation of authors, and domain. Figure ~\ref{fig:2figsA} shows the number of primary articles published each year. The first article was from 2005, and from 2006-2008 no articles were found. From 2009 until 2011, we identified one article per year. In 2019, the highest number of publications (N=8) was found. After 2019, there was a decrease in the number of publications per year; however, these numbers (four publications per year) were still higher than from 2005 until 2018. Figure ~\ref{fig:2figsB} shows the proportions of articles by publication type. We found that most publications were conference publications (41.2\% of all the publications). Workshop publications made up 26.5\%, journal publications 20.6\%, and book chapters and symposium publications 5.9\% each. The most frequent conference venue for publications was the International Requirements Engineering Conference (P1, P6, P29), while the top workshop venues were the International Workshop on Affective Computing for Requirements Engineering  (P25, P30, P22), and the International Workshop on Requirements Engineering for Well-Being, Aging, and Health  (P31, 33). The top journal venue was the Journal of Systems and Software (P3, P4).

 Our selected studies included 88 unique authors, 75 authors had purely academic affiliations, while 13 authors were affiliated with industry i.e. research centres, hospitals, or companies. Purely academic authors were affiliated with 30 different universities, among which the highest number of researchers was affiliated with the University of Melbourne (seven authors) and Deakin University (six authors) from Australia, the most active authors among the selected publications are listed in Table ~\ref{tab: Most active researchers and their affiliations}. With regards to the geographical location of the authors shown in Figure \ref{fig:authorlocations}, Australia is the leading country (18 authors) closely followed by UK (14 authors), Spain (11 authors), and Malaysia (11 authors). 

\input{Tables/author}

Using the data extraction method described in Section ~\ref{sec:Extraction}, domain categories were identified. The identified domain categories were smart homes, e-learning, well-being, healthcare, mobile applications, information systems, game development, and automated vehicles. 
Table ~\ref{tab: Domain_Categories} shows the distribution of the selected publications across domain categories. Out of the 34 selected publications, ten publications belonged to the healthcare domain, while seven publications were concerned with the domain of well-being. Six publications addressed improving the RE process in the domain of information systems. The domains of mobile application development, game development, e-learning, and smart homes were respectively explored by four, two, two, and three publications. The domain of automated vehicles was the least explored domain - it was treated by just one publication. Only one study (P4) was associated with two domains at the same time, including smart home and well-being.

\input{Tables/Domain_Categories}

\subsection{\textbf{RQ2.} How are emotions elicited and represented in different phases of RE?}
\label{sec:4.2}

To answer RQ2, we classified the selected 34 publications according to the instrument or technique they employed for eliciting emotional requirements. The instrument was identified using the data extraction method described in Section ~\ref{sec:Extraction}. The instruments used in the RE process to elicit emotional requirements are listed in Table \ref{tab: instrument_to _elicit}. Eight studies (P3, P4, P7, P11, P16, P19, P24, P33) reported on using multiple techniques for eliciting emotional requirements. Seventeen publications reported the interview technique as an elicitation instrument. Different interview techniques were used, such as  semi-structured interviews (P1, P6, P22, P24), structured interviews (P30), and a combination of both structured and semi-structured interviews (P11). The remaining 11 publications (P3, P4, P9, P12, P15, P16, P17, P19, P26, P31, P33) did not specify the type of interview conducted. Eight publications (P3, P4, P7, P8, P16, P23, P33, P34) reported on using a questionnaire/survey, and five publications (P9, P17, P18, P20, P30) on using the workshop as an elicitation instrument. Observation, feedback, and personas/scenarios were reported by three publications each. Two publications per each instrument category used focus groups (P11, P24), literature review (P19, P27) and document analysis (P4, P7).

\input{Tables/Instrument_to_Elicit}

In our selected publications, two publications (P8, P17) focused specifically on the requirements representation phase of RE. One publication (P14) was concerned with the improvement of the RE process in general. In total, above 90\% of the publications (31 out of 34) focused on the requirements elicitation phase of RE. Elicitation was exclusively addressed by 13 publications (P3, P4, P10, P12, P18, P19, P20, P21, P26, P27, P30, P31, P34). The remaining publications were concerned with a combination of RE phases such as elicitation and representation (P6, P9, P11, P15), elicitation and analysis (P16, P24, P32, P33), elicitation and prioritization (P1, P23), elicitation and validation (P5, P13), and elicitation and evaluation (P2, P7, P22, P25, P28, P29).

\subsection{\textbf{RQ3.} What is the state of the practice in emotion-related RE?} 
\label{sec:4.3}

To answer RQ3, we extracted data from the selected publications to identify what kinds of artefacts were produced to aid the process of engineering emotional requirements. Regarding tools, only one publication (P12) reported on having developed a tool that practitioners could use to improve the RE process, including engineering emotional requirements. The identified tool is a web-based solution for assisting analysts in preparing for stakeholder meetings. However, this solution is neither open-source nor publicly accessible. In terms of approaches, most of the selected publications explained how they utilized or modified the existing artefacts, such as models, frameworks, or guidelines, 
for eliciting and/or representing emotional requirements. 
\input{Tables/Artifacts_generated_modified}

Table \ref{tab: Artifacts_generated_modified} shows details of utilized or modified artefacts reported by the selected publications. The studies P17, P18, P27, P32, and P33 reported on producing artefacts of more than one type. We identified various types of artefacts ranging from frameworks, user stories, and personas to taxonomies and guidelines. Utilizing or modifying models of the existing types that could aid the process of engineering emotional requirements was reported by 60\% of the studies (N = 18). In the column `Primary publication(s)' of Table \ref{tab: Artifacts_generated_modified}, we further specify the type of the model explored by the corresponding publication. Out of the 18 publications that reported on producing models, eleven publications (P3, P4, P8, P15, P18, P19, P22, P24, P27, P28, P30) focused on motivational goal models. Two publications (P13, P17) reported on using agent storyboards and presented models and processes for analyzing the role of emotions in interactive user-centred applications. Models for mapping emotions from user data were reported in two publications (P1, P25). Three respective publications reported agent models (P27), generative models (P26), and data mining and statistical models (P34). Four publications (P11, P12, P16, P17) produced guidelines that could be used in value-based requirements elicitation and preliminary design of software systems. Other artefacts reported in the publications were customer journey maps (P31), user stories (P18, P33), use case scenarios (P18), personas (P10, P33), case studies (P20), affect grids (P2, P14), domain-specific modelling language (P5), list of features from user reviews (P29), multi-layered scenario analysis method (P23), hybrid Kansei clustering method (P7), dialogue fragments (P21), and paper-based and digital drawings (P9).

%% file: Tables/author.tex
\begin{table}[!b]
\centering
\caption{Most active researchers in selected publications}
\label{tab: Most active researchers and their affiliations}
\begin{tabularx}{.48\textwidth}{X|X|p{3.5cm}}

\hline
\textbf{Author} &	\textbf{Ref.} &	\textbf{Affiliation}  \\ \hline
Tim Miller &	P3, P6, P8, P10, P15, P24 &	University of Melbourne, Australia \\ \hline
Sonja Pedell &	P3, P6, P10, P15, P30 &	Swinburne University of Technology, Australia \\ \hline
Antonette Mendoza &	P3, P6, P8, P10, P24 &	University of Melbourne, Australia \\ \hline
Leon Sterling &	P3, P6, P10, P15, P30 &	Swinburne University of Technology, Australia \\ \hline
John Grundy &	P4, P28, P33 &	Deakin University, Australia \& Monash University, Australia \\ \hline
Rachel Burrows &	P6, P8, P30 &	University of Melbourne, Australia \& Cambridge Science Park, UK \\ \hline
Alistair Sutcliffe &	P12, P13, P17 &	University of Manchester, UK \\ \hline
Kuldar Taveter &	P18, P19, P30 &	University of Tartu, Estonia \\ \hline
Ricardo Colomo-Palacios&P2, P14&	Universidad Carlos III de Madrid, Spain	\\ \hline
Simon André Scherr&	P1, P25	&Fraunhofer IESE, Kaiserslautern, Germany	\\ \hline
Patrick Mennig	&P1, P25 &	Fraunhofer IESE, Kaiserslautern, Germany	\\ \hline
Pedro Soto-Acosta&	P2, P14	&University of Murcia,	Spain\\ \hline
Ángel García-Crespo&	P2, P14&	Universidad Carlos III de Madrid, Spain	\\ \hline
Frank Elberzhager&	P1 , P25 &	University of Applied Sciences and Arts Northwestern, Switzerland \& Fraunhofer IESE, Kaiserslautern	Germany	\\ \hline
Alen Keirnan &P3, P10 &	Swinburne University of Technology,	Australia\\ \hline
Antonio A. Lopez-Lorca &	P3, P10 &	Swinburne University of Technology, Australia	\\ \hline
\end{tabularx}

\end{table}

%% file: Tables/Domain_Categories.tex
\begin{table}[!t]
\centering
\caption{Distribution of selected publications across domain categories}
\label{tab: Domain_Categories}
\begin{tabularx}{.48\textwidth}{p{2.6cm}|X|c}

\hline
\textbf{Domain} & \textbf{Primary publication(s)} & \textbf{\#} \\ \hline

Healthcare & P11, P13, P16, P18, P19, P20, P22, P24, P30, P33 & 10  \\ \hline
Well-being & P4, P6, P9, P10, P15, P17, P31 & 7 \\ \hline
Information systems &  P2, P7, P12, P14, P21, P26 & 6 \\ \hline
Mobile applications & P1, P25, P29, P34 & 4 \\ \hline
Smart home & P3, P4, P28  & 3 \\ \hline
E-learning & P8, P27 & 2 \\ \hline
Game development & P5, P32 & 2 \\ \hline
Automated vehicles & P23 & 1 \\ \hline
\end{tabularx}

\end{table}

%% file: Tables/Instrument_to_Elicit.tex
\begin{table}[!ht]
\centering
\caption{Instruments used to elicit emotional requirements}
\label{tab: instrument_to _elicit}
\begin{tabularx}{.48\textwidth}{p{3cm}|p{4cm}|c}

\hline
\textbf{Instrument} & \textbf{Primary publication(s)} &\textbf{\# } \\ \hline
Interview & P1, P3, P4, P6, P9, P11, P12, P15, P16, P17, P19, P22, P24, P26, P30, P31, P33 & 17\\ \hline
Questionnaire / survey & P3, P4, P7, P8, P16, P23, P33, P34 & 8  \\ \hline
Workshop &  P9, P17, P18, P20, P30 & 5 \\ \hline
Observation & P11, P21, P33 & 3 \\ \hline
Feedback & P7, P15, P29 & 3 \\ \hline
Personas / scenarios &  P5, P10, P32 &3  \\ \hline
Document analysis & P4, P7 & 2\\ \hline
Focus group & P11, P24  & 2\\ \hline
Literature review & P19, P27 & 2 \\ \hline
Sensor/physiological data & P25 & 1 \\ \hline

\end{tabularx}

\end{table}

%% file: Tables/Artifacts_generated_modified.tex
\begin{table}[!ht]
\centering
\caption{Artifacts generated or modified for eliciting and representing emotional requirements}
\label{tab: Artifacts_generated_modified}
\begin{tabularx}{.48\textwidth}{p{3cm}|p{4.2cm}|p{0.4cm}}

\hline
\textbf{Artifacts}                  & \textbf{Primary publication(s)} & \textbf{\#} \\ \hline
\multirow{2}{*}{Guidelines} & \textit{Value-based elicitation guidelines:} P12 &  1\\ \cline{2-3}
                                    & \textit{Design guidelines:} P11, P16, P17                     & 3 \\ \hline

 \multirow{7}{*}{Models} 
                                    & \textit{Agent storyboards:} P13, P17                & 2\\\cline{2-3}
                                    & \textit{Models for mapping emotions from usage data:} P1, P25        & 2\\ \cline{2-3}
                                    & \textit{Generative models:} P26                                & 1\\ \cline{2-3}
                                    & \textit{Motivational goal models: } P3, P4, P8, P15, P18, P19, P22, P24, P27, P28, P30                     & 11\\ \cline{2-3}
                                    & \textit{Statistical models:} P34             & 1 \\ \cline{2-3}
                                    & \textit{Agent models:} P27            & 1\\ \hline
Customer journey maps                & P31                                                           & 1\\ \hline
User stories                          & P18, P33                                                      & 2\\ \hline
Use case scenarios                   & P18                                                           & 1\\ \hline
Personas                            & P10, P33                                                      & 2\\ \hline
Case study                          & P20                                                           & 1 \\ \hline
Affect grids                         & P2, P14                                                       & 2\\ \hline
Domain-specific modelling languages  & P5                                                        
  & 1\\ \hline 
Features                           & P29                                                        
  & 1\\ \hline
Multi-layered scenario analysis method & P23                                                        & 1\\ \hline
Hybrid Kansei clustering method     & P7                                                            & 1 \\ \hline
Dialogue fragments                   & P21                                                           & 1\\ \hline
Paper-based and digital drawings & P9                                       & 1\\ \hline
Emoticons, emotion intensity maps, emotion timelines & P32                                       & 1\\ \hline
\end{tabularx}%

\end{table}

%% file: Disscussion.tex
\section{Discussion}
\label{sec:Discussion}



In this section, we will discuss the answers provided in Section \ref{sec:results} to the research questions formulated in Section ~\ref{sec:research_questions} in more detail. Furthermore, we will also discuss threats to the validity of our study. 

With respect to the research question RQ1, emotional requirements have gained more visibility in the RE community, which can be noted as an increase in the number of relevant publications according to the study conducted by us. Between 2009 and 2021, at least one article on emotional requirements has been published every year. The highest number of publications (8 publications) appeared in the year 2019 and the number of publications has stabilized after that at a higher level (4 publications per year). This may indicate an increase in the recognition of the importance of emotional requirements.

More specifically, our mapping study has identified several essential application domains where emotional requirements have been considered. Among them, healthcare and human well-being were the two most explored domains in the selected publications. This can be explained by the fact that staying healthy and well-being are important human values of the sub-type \textit{personal security} ~\cite{schwartz2017refined} and people feel emotionally aroused when their values are threatened \cite{schwartz2012overview}. Several analysed publications concerned with healthcare or well-being focused on capturing the emotions of older adults, children, and patients in an effort to improve human-computer interaction. On the other hand, the domains of smart homes, game development, e-learning, and automated vehicles were not very well explored in engineering emotional requirements. This is surprising because these are all emotion-intense application domains. This may be due to the fact that in these domains, RE is usually not explicitly performed, which is also reflected by a relatively small number of relevant publications. Therefore, the field of emotions in RE should focus more on case studies in the above-mentioned domains.

For the research question RQ2, which focuses on how emotions are elicited and represented in different
phases of RE, our study has identified various instruments for eliciting emotional requirements. Among them, the interview technique was the most frequently used instrument for eliciting emotional requirements. The interview technique was followed by questionnaires/surveys and workshops. The relatively high usage of questionnaires/surveys for the elicitation of emotional requirements is surprising because quantitative methods generally do not lend themselves well to capturing emotional aspects. On the other side of the spectrum, the use of sensor data / physiological data was the least frequently used instrument for eliciting emotional requirements. However, sensor data and physiological data have been used for recognizing emotions by software developers to identify the impact of emotions on their productivity \cite{fountaine2017emotional, muller2015stuck}. They have also been used in autonomous vehicles for the classification of the emotional state of passengers \cite{park2018eeg}.

With respect to the research question RQ3, we have identified and discussed various artefacts that were generated or modified for eliciting and representing emotional requirements. However, we could not identify a clear trend in the types of artefacts produced. This could be due to the fact that emotional requirements have yet to receive a lot of attention from the research community, and therefore most of the work done in this area is still experimental. We also noticed a broad usage of different kinds of models, such as goal models, generative models, and data mining and statistical models. It is also noteworthy that only one publication (P12) reported on a proprietary tool that practitioners could use for engineering emotional requirements. Most of the artefacts were used in the requirements elicitation and representation phases which indicates the need to increasingly target research efforts at the validation phase of engineering emotional requirements.

We emphasize presenting the results of the study in such a manner that they can be easily understood and applied. This may encourage software engineers to consider emotional requirements during RE and system design and facilitate dealing with such requirements. Additionally, our results may reduce the vagueness when dealing with emotional requirements by providing a better understanding of what they are, the context in which they can be used, and the benefits of considering them during RE and system design. This will also help both stakeholders and software engineers to communicate better and understand each other, which may prevent wrong design decisions based on the requirements. Our study is novel; to the best of the knowledge of the authors, no other similar mapping study has been conducted on emotional requirements.

The methods for engineering emotional requirements should be evaluated in real software development environments for usability and usefulness. Considering this, the proposed methods should be checked for scalability in an industrial setting with longer interviews and larger datasets. Such methods should also ensure that the once elicited and represented emotional requirements are also manifested in the system design and are not "lost" in the design process.

%% file: ThreatsToValidity.tex
\subsection{Threats to Validity}
\label{sec:ThreatstoValidity}

Following Runeson, et al. \cite{runeson2009guidelines}, we address threats to validity under the following categories:

\noindent  \textbf{1- Internal threats} focus on factors that have not been considered in the study but may have influenced the results. We have identified the following internal threats: (i) \textit{Systematic error} is a common threat to the validity of a mapping study and can occur during the study design and/or conducting the systematic mapping studies. To mitigate this kind of threat, the study protocol has been carefully designed based on well-adopted methods, and the protocol has been strictly followed during different phases of the study. (ii) \textit{Authors' background}. Three authors of this study are working and have experience with emotional requirements, which may influence their opinion towards the selection of publications with which they are more familiar. However, the other three authors do not work in the area of emotional requirements. To mitigate the threat of favouring the selection of some publications based on familiarity, we have formed teams of two people, where one of the team members is working on emotional requirements and the other is not. 


\noindent  \textbf{2-  External threats} focus on the extent to which the results of the study can be generalized. We have identified the following external threats: (i) \textit{Completeness of relevant publications}. It is almost impossible to guarantee that a mapping study can identify all relevant publications, yet we did our best to mitigate this threat as our mapping study protocol and search strategy were very carefully designed to retrieve as many relevant publications as possible. (ii) \textit{Limited number of selected publications.} We have selected 34 publications, which might be considered as a limited number to produce concrete results and draw valid conclusions. However, we should consider that emotions in requirements engineering is a relatively new research area. Moreover, our inclusion/exclusion criteria, as well as our quality assessment criteria, were carefully formulated to identify relevant publications and reduce the noise by excluding non-relevant publications. This also explains why we were not able to identify more than 34 publications.

\noindent  \textbf{3- Construct threats}. The primary threat to construct validity occurs in the data extraction phase. To mitigate the construct threat, a trial session was conducted with all of the involved researchers to align their understanding of the research questions and data to be extracted. Also, after the full extraction run, the results were discussed and classified into different categories with mutual agreements, and only a few changes were made.

\noindent  \textbf{4- Reliability threats} focus on the extent to which the study is dependent on the researcher(s), i.e., when the study is repeated by other researchers, the results should be comparable. Our study, including determining sources to be searched, inclusion/exclusion criteria, and quality assessment criteria, are all well documented and made available. To this end, if other researchers would repeat the study, they should get comparable results.  

%% file: Conclusion.tex
\section{Conclusions}
\label{sec:conclusion}
The study conducted by us has identified 34 publications that address a wide spectrum of practices concerned with emotional requirements. In our paper, these publications have been further analyzed to answer our research questions and derive future research directions that will benefit engineering emotional requirements. We are confident that our results reduce the vagueness when dealing with emotional requirements by providing a better understanding of what they are, the context in which they can be used, and the benefits of considering them during RE and system design. This will also help both stakeholders and software engineers to communicate better and understand each other, which may prevent wrong design decisions based on the requirements. Based on our mapping study, we can also conclude that engineering emotional requirements is an interdisciplinary effort that involves software engineers, product designers, psychologists, and end users.

Our mapping study reveals that there is a need to increasingly target research efforts at the validation phase of engineering emotional requirements. Since emotions are strongly impacted by cultural factors, we also identified the need for artefacts geared towards dealing with cross-cultural scenarios and addressing the RE process in the global environment. There is also a need for visual design environments for engineering emotional requirements.



%% file: Tables/ref_table.tex

\footnotesize{
\centering
\begin{supertabular}{p{.21in}|p{.11in}|p{5.55in}}
\label{tab:ref_table}
\tablefirsthead
\hline

\textbf{Ref} & \textbf{ID} & \textbf{Title} \\ 
\hline

\cite{P1}  & P1  & Don't Worry, Be Happy – Exploring Users' Emotions During App Usage for Requirements Engineering    \\ \hline
\cite{P2}  & P2  & Using the affect grid to measure emotions in software requirements engineering                 \\ \hline
\cite{P3}  & P3  & Emotion-led modelling for people-oriented requirements engineering: The case study of emergency systems  \\ \hline
\cite{P4}  & P4  & Emotion-oriented requirements engineering: A case study in developing a smart home system for the elderly  \\ \hline
 \cite{P5}  & P5  & Towards a requirements language for modeling emotion in videogames                                   \\ \hline
\cite{P6}  & P6  & Motivational modelling in Software for Homelessness: Lessons from an Industrial Study                 \\ \hline
\cite{P7}  & P7  & Consumer-Oriented Emotional Design Using a Correlation Handling Strategy                              \\ \hline
\cite{P8}    & P8  & Emotionalism Within People-Oriented Software Design                                               \\ \hline
\cite{P9}  & P9  & Characters as Agents for the Co-Design Process                                                       \\ \hline
\cite{P10} & P10 & One Size Doesn't Fit All: Diversifying the User Using Personas and Emotional Scenarios           \\\hline
\cite{P11} & P11 & Assistive technology for older adults: Psychological and socio-emotional design requirements \\\hline
\cite{P12} & P12 & Value-based requirements engineering: method and experience \\\hline
\cite{P13} & P13 & User-Oriented Requirements Engineering  \\\hline
\cite{P14} & P14 & A Study of Emotions in Requirement Engineering  \\\hline
\cite{P15} & P15 & Substantiating Agent-Based Quality Goals for Understanding Socio-Technical Systems   \\\hline
\cite{P16} & P16 & Research on the Design of Mobile Infusion Devices for Children Based on Emotionalization \\\hline
\cite{P17} & P17 & Designing User Interfaces in Emotionally-Sensitive Applications \\\hline
\cite{P18} & P18 & Co-design and engineering of user requirements for a novel ICT healthcare solution in Murcia, Spain \\\hline
\cite{P19} & P19 & Agent-Oriented Goal Models in Developing Information Systems Supporting Physical Activity Among Adolescents: Literature Review and Expert Interviews \\\hline
\cite{P20} & P20 & Developing mHealth Apps with Researchers: Multi-Stakeholder Design Considerations \\\hline
\cite{P21} & P21 & Older adults: Key factors in design  \\\hline
\cite{P22} & P22 & Design of a Remote Emotional Requirement Elicitation Feedback Method        \\\hline
\cite{P23} & P23 & MuLSA: Multi-layered scenario analysis for an advanced driver assistance system \\\hline
\cite{P24} & P24 & Psychologically-Driven Requirements Engineering: A Case Study in Depression Care \\\hline
\cite{P25} & P25 & On the Road to Enriching the App Improvement Process with Emotions 
\\\hline
\cite{P26} & P26 & ELICA: An Automated Tool for Dynamic Extraction of Requirements Relevant Information 
\\\hline
\cite{P35} & P27 & Modelling emotion expression through agent oriented methodology 
\\\hline
\cite{P28} & P28 & Human-centric software engineering for next generation cloud-and edge-based smart living applications 
\\\hline
\cite{P29} & P29 & How do users like this feature? a fine grained sentiment analysis of app reviews 
\\\hline
\cite{P30} & P30 & A method for eliciting and representing emotional requirements: Two case studies in e-healthcare 
\\\hline
\cite{P31} & P31 & Emotional requirements for well-being applications: The customer journey 
\\\hline
\cite{P32} & P32 & Visualizing Emotional Requirements 
\\\hline
\cite{P33} & P33 & Using Work System Design, User Stories and Emotional Goal Modeling for an mHealth System 
\\\hline
\cite{P34} & P34 & A Holistic Approach to Requirements Elicitation for Mobile Tourist Recommendation Systems  
\\\hline


\end{supertabular}
}


%% file: CHASE.bbl
\begin{thebibliography}{10}
\providecommand{\url}[1]{#1}
\csname url@samestyle\endcsname
\providecommand{\newblock}{\relax}
\providecommand{\bibinfo}[2]{#2}
\providecommand{\BIBentrySTDinterwordspacing}{\spaceskip=0pt\relax}
\providecommand{\BIBentryALTinterwordstretchfactor}{4}
\providecommand{\BIBentryALTinterwordspacing}{\spaceskip=\fontdimen2\font plus
\BIBentryALTinterwordstretchfactor\fontdimen3\font minus
  \fontdimen4\font\relax}
\providecommand{\BIBforeignlanguage}[2]{{%
\expandafter\ifx\csname l@#1\endcsname\relax
\typeout{** WARNING: IEEEtran.bst: No hyphenation pattern has been}%
\typeout{** loaded for the language `#1'. Using the pattern for}%
\typeout{** the default language instead.}%
\else
\language=\csname l@#1\endcsname
\fi
#2}}
\providecommand{\BIBdecl}{\relax}
\BIBdecl

\bibitem{Softwareengineering8}
I.~Sommerville, \emph{Engineering software products}.\hskip 1em plus 0.5em
  minus 0.4em\relax Pearson London, 2020.

\bibitem{chung2009non}
L.~Chung and J.~C. S.~d. Prado~Leite, ``On non-functional requirements in
  software engineering,'' in \emph{Conceptual modeling: Foundations and
  applications}.\hskip 1em plus 0.5em minus 0.4em\relax Springer, 2009, pp.
  363--379.

\bibitem{chung2012non}
L.~Chung, B.~A. Nixon, E.~Yu, and J.~Mylopoulos, \emph{Non-functional
  requirements in software engineering}.\hskip 1em plus 0.5em minus 0.4em\relax
  Springer Science \& Business Media, 2012, vol.~5.

\bibitem{curumsing2017emotion}
M.~K. Curumsing, ``Emotion-oriented requirements engineering,'' \emph{Ph. D.
  dissertation}, 2017.

\bibitem{Taveter}
K.~Taveter and T.~Iqbal, ``Theory of constructed emotion meets re,'' in
  \emph{2021 IEEE 29th International Requirements Engineering Conference
  Workshops (REW)}, 2021, pp. 383--386.

\bibitem{P19}
K.~Mooses and K.~Taveter, ``Agent-oriented goal models in developing
  information systems supporting physical activity among adolescents:
  Literature review and expert interviews,'' \emph{Journal of Medical Internet
  Research}, vol.~23, 2021.

\bibitem{P30}
K.~Taveter, L.~Sterling, S.~Pedell, R.~Burrows, and E.~M. Taveter, ``A method
  for eliciting and representing emotional requirements: Two case studies in
  e-healthcare,'' in \emph{2019 IEEE 27th International Requirements
  Engineering Conference Workshops (REW)}.\hskip 1em plus 0.5em minus
  0.4em\relax IEEE, 2019, pp. 100--105.

\bibitem{mendoza2013role}
A.~Mendoza, T.~Miller, S.~Pedell, and L.~Sterling, ``The role of users’
  emotions and associated quality goals on appropriation of systems: Two case
  studies,'' in \emph{24th Australasian Conference on Information Systems},
  2013.

\bibitem{ramos2005emotion}
I.~Ramos and D.~M. Berry, ``Is emotion relevant to requirements engineering?''
  \emph{Requirements Engineering}, vol.~10, no.~3, pp. 238--242, 2005.

\bibitem{pedellemotions}
S.~Pedell, L.~Sterling, H.~Davis, A.~Keirnan, and G.~Dobson, \emph{Emotions
  around emergency alarm use: A field study with older adults. Report for Smart
  Services CRC Personalisation Project H5}.\hskip 1em plus 0.5em minus
  0.4em\relax Swinburne University of Technology, 2013.

\bibitem{miller2015emotion}
T.~Miller, S.~Pedell, A.~A. Lopez-Lorca, A.~Mendoza, L.~Sterling, and
  A.~Keirnan, ``Emotion-led modelling for people-oriented requirements
  engineering: the case study of emergency systems,'' \emph{Journal of Systems
  and Software}, vol. 105, pp. 54--71, 2015.

\bibitem{iqbal2023theory}
T.~Iqbal, J.~G. Marshall, K.~Taveter, and A.~Schmidt, ``Theory of constructed
  emotion meets re: An industrial case study,'' \emph{Journal of Systems and
  Software}, vol. 197, p. 111544, 2023.

\bibitem{P4}
M.~K. Curumsing, N.~Fernando, M.~Abdelrazek, R.~Vasa, K.~Mouzakis, and
  J.~Grundy, ``Emotion-oriented requirements engineering: A case study in
  developing a smart home system for the elderly,'' \emph{Journal of systems
  and software}, vol. 147, pp. 215--229, 2019.

\bibitem{P8}
\BIBentryALTinterwordspacing
M.~Sherkat, T.~Miller, A.~Mendoza, and R.~Burrows, ``Emotionalism within
  people-oriented software design,'' \emph{Frontiers in Computer Science},
  vol.~3, 2021. [Online]. Available:
  \url{https://www.frontiersin.org/articles/10.3389/fcomp.2021.717787}
\BIBentrySTDinterwordspacing

\bibitem{GeneratingRequirementsOutofThinAir}
T.~Iqbal, N.~Seyff, and D.~Mendez, ``Generating requirements out of thin air:
  Towards automated feature identification for new apps,'' in \emph{2019 IEEE
  27th International Requirements Engineering Conference Workshops (REW)},
  2019, pp. 193--199.

\bibitem{lopez2014modelling}
A.~A. Lopez-Lorca, T.~Miller, S.~Pedell, L.~Sterling, and M.~K. Curumsing,
  ``Modelling emotional requirements,'' 2014.

\bibitem{curumsing2019emotion}
M.~K. Curumsing, N.~Fernando, M.~Abdelrazek, R.~Vasa, K.~Mouzakis, and
  J.~Grundy, ``Emotion-oriented requirements engineering: A case study in
  developing a smart home system for the elderly,'' \emph{Journal of systems
  and software}, vol. 147, pp. 215--229, 2019.

\bibitem{petersen2008systematic}
K.~Petersen, R.~Feldt, S.~Mujtaba, and M.~Mattsson, ``Systematic mapping
  studies in software engineering,'' in \emph{12th International Conference on
  Evaluation and Assessment in Software Engineering (EASE) 12}, 2008, pp.
  1--10.

\bibitem{kitchenham2007cross}
B.~A. Kitchenham, E.~Mendes, and G.~H. Travassos, ``Cross versus within-company
  cost estimation studies: A systematic review,'' \emph{IEEE Transactions on
  Software Engineering}, vol.~33, no.~5, pp. 316--329, 2007.

\bibitem{kitchenham2007guidelines}
B.~Kitchenham and S.~Charters, ``Guidelines for performing systematic
  literature reviews in software engineering,'' 2007.

\bibitem{petersen2015guidelines}
K.~Petersen, S.~Vakkalanka, and L.~Kuzniarz, ``Guidelines for conducting
  systematic mapping studies in software engineering: An update,''
  \emph{Information and software technology}, vol.~64, pp. 1--18, 2015.

\bibitem{zhao2021natural}
L.~Zhao, W.~Alhoshan, A.~Ferrari, K.~J. Letsholo, M.~A. Ajagbe, E.-V. Chioasca,
  and R.~T. Batista-Navarro, ``Natural language processing for requirements
  engineering: a systematic mapping study,'' \emph{ACM Computing Surveys
  (CSUR)}, vol.~54, no.~3, pp. 1--41, 2021.

\bibitem{schwartz2017refined}
S.~H. Schwartz, ``The refined theory of basic values,'' in \emph{Values and
  behavior}.\hskip 1em plus 0.5em minus 0.4em\relax Springer, 2017, pp. 51--72.

\bibitem{schwartz2012overview}
S.~H. Schwartz \emph{et~al.}, ``An overview of the schwartz theory of basic
  values,'' \emph{Online readings in Psychology and Culture}, vol.~2, no.~1,
  pp. 2307--0919, 2012.

\bibitem{fountaine2017emotional}
A.~Fountaine and B.~Sharif, ``Emotional awareness in software development:
  Theory and measurement,'' in \emph{2017 IEEE/ACM 2nd International Workshop
  on Emotion Awareness in Software Engineering (SEmotion)}.\hskip 1em plus
  0.5em minus 0.4em\relax IEEE, 2017, pp. 28--31.

\bibitem{muller2015stuck}
S.~C. M{\"u}ller and T.~Fritz, ``Stuck and frustrated or in flow and happy:
  Sensing developers' emotions and progress,'' in \emph{2015 IEEE/ACM 37th IEEE
  International Conference on Software Engineering}, vol.~1.\hskip 1em plus
  0.5em minus 0.4em\relax IEEE, 2015, pp. 688--699.

\bibitem{park2018eeg}
C.~Park, S.~Shahrdar, and M.~Nojoumian, ``Eeg-based classification of emotional
  state using an autonomous vehicle simulator,'' in \emph{2018 IEEE 10th Sensor
  Array and Multichannel Signal Processing Workshop (SAM)}.\hskip 1em plus
  0.5em minus 0.4em\relax IEEE, 2018, pp. 297--300.

\bibitem{runeson2009guidelines}
P.~Runeson and M.~H{\"o}st, ``Guidelines for conducting and reporting case
  study research in software engineering,'' \emph{Empirical software
  engineering}, vol.~14, no.~2, pp. 131--164, 2009.

\bibitem{P1}
M.~Stade, S.~A. Scherr, P.~Mennig, F.~Elberzhager, and N.~Seyff, ``Don't worry,
  be happy – exploring users' emotions during app usage for requirements
  engineering,'' in \emph{2019 IEEE 27th International Requirements Engineering
  Conference (RE)}, 2019, pp. 375--380.

\bibitem{P2}
R.~Colomo-Palacios, C.~Casado-Lumbreras, P.~Soto-Acosta, and
  {\'A}.~Garc{\'\i}a-Crespo, ``Using the affect grid to measure emotions in
  software requirements engineering,'' \emph{Journal of Universal Computer
  Science}, pp. 1281--1298, 2011.

\bibitem{P3}
T.~Miller, S.~Pedell, A.~A. Lopez-Lorca, A.~Mendoza, L.~Sterling, and
  A.~Keirnan, ``Emotion-led modelling for people-oriented requirements
  engineering: the case study of emergency systems,'' \emph{Journal of Systems
  and Software}, vol. 105, pp. 54--71, 2015.

\bibitem{P5}
G.~Migu{\'e}is, J.~Araujo, and A.~Moreira, ``Towards a requirements language
  for modeling emotion in videogames,'' in \emph{Proceedings of the 34th
  ACM/SIGAPP Symposium on Applied Computing}, 2019, pp. 1878--1880.

\bibitem{P6}
R.~Burrows, A.~Lopez-Lorca, L.~Sterling, T.~Miller, A.~Mendoza, and S.~Pedell,
  ``Motivational modelling in software for homelessness: Lessons from an
  industrial study,'' in \emph{2019 IEEE 27th International Requirements
  Engineering Conference (RE)}, 2019, pp. 297--307.

\bibitem{P7}
D.~Chang, Y.~Huang, C.-H. Chen, and L.~P. Khoo, ``Consumer-oriented emotional
  design using a correlation handling strategy.'' in \emph{ISPE CE}, 2015, pp.
  184--193.

\bibitem{P9}
\BIBentryALTinterwordspacing
C.~Grundy, L.~Pemberton, and R.~Morris, ``Characters as agents for the
  co-design process,'' in \emph{Proceedings of the 11th International
  Conference on Interaction Design and Children}, ser. IDC '12.\hskip 1em plus
  0.5em minus 0.4em\relax New York, NY, USA: Association for Computing
  Machinery, 2012, p. 180–183. [Online]. Available:
  \url{https://doi.org/10.1145/2307096.2307120}
\BIBentrySTDinterwordspacing

\bibitem{P10}
A.~A. Lopez-Lorca, T.~Miller, S.~Pedell, A.~Mendoza, A.~Keirnan, and
  L.~Sterling, ``One size doesn't fit all: Diversifying `the user' using
  personas and emotional scenarios,'' in \emph{Proceedings of the 6th
  International Workshop on Social Software Engineering}, 2014, pp. 25--32.

\bibitem{P11}
A.~K. Bright and L.~Coventry, ``Assistive technology for older adults:
  psychological and socio-emotional design requirements,'' in \emph{Proceedings
  of the 6th international conference on pervasive technologies related to
  assistive environments}, 2013, pp. 1--4.

\bibitem{P12}
S.~Thew and A.~Sutcliffe, ``Value-based requirements engineering: method and
  experience,'' \emph{Requirements Engineering}, vol.~23, no.~4, pp. 443--464,
  2018.

\bibitem{P13}
A.~Sutcliffe, ``User-oriented requirements engineering,'' in
  \emph{Usability-and Accessibility-Focused Requirements Engineering}.\hskip
  1em plus 0.5em minus 0.4em\relax Springer, 2012, pp. 11--33.

\bibitem{P14}
R.~Colomo-Palacios, A.~Hern{\'a}ndez-L{\'o}pez, {\'A}.~Garc{\'\i}a-Crespo, and
  P.~Soto-Acosta, ``A study of emotions in requirements engineering,'' in
  \emph{World Summit on Knowledge Society}.\hskip 1em plus 0.5em minus
  0.4em\relax Springer, 2010, pp. 1--7.

\bibitem{P15}
S.~Pedell, T.~Miller, L.~Sterling, F.~Vetere, and S.~Howard, ``Substantiating
  agent-based quality goals for understanding socio-technical systems,'' in
  \emph{International Conference on Autonomous Agents and Multiagent
  Systems}.\hskip 1em plus 0.5em minus 0.4em\relax Springer, 2011, pp. 80--95.

\bibitem{P16}
X.~Zhao, W.~Yu, and X.~Liang, ``Research on the design of mobile infusion
  devices for children based on emotionalization,'' in \emph{International
  Conference on Human-Computer Interaction}.\hskip 1em plus 0.5em minus
  0.4em\relax Springer, 2021, pp. 170--185.

\bibitem{P17}
A.~Sutcliffe, ``Designing user interfaces in emotionally-sensitive
  applications,'' in \emph{IFIP Conference on Human-Computer
  Interaction}.\hskip 1em plus 0.5em minus 0.4em\relax Springer, 2017, pp.
  404--422.

\bibitem{P18}
R.~Mart{\'\i}nez, F.~J. Moreno-Muro, F.~J. Melero-Mu{\~n}oz, M.~V.
  Bueno-Delgado, J.~Garrido-Lova, M.~S{\'a}nchez-Melero, and K.~Taveter,
  ``Co-design and engineering of user requirements for a novel ict healthcare
  solution in murcia, spain,'' in \emph{International Conference on Smart
  Objects and Technologies for Social Good}.\hskip 1em plus 0.5em minus
  0.4em\relax Springer, 2021, pp. 279--292.

\bibitem{P20}
M.~P. Craven, A.~R. Lang, and J.~L. Martin, ``Developing mhealth apps with
  researchers: multi-stakeholder design considerations,'' in
  \emph{International Conference of Design, User Experience, and
  Usability}.\hskip 1em plus 0.5em minus 0.4em\relax Springer, 2014, pp.
  15--24.

\bibitem{P21}
M.~Zajicek, ``Older adults: Key factors in design,'' in \emph{Future
  interaction design}.\hskip 1em plus 0.5em minus 0.4em\relax Springer, 2005,
  pp. 151--176.

\bibitem{P22}
E.~Jackson and A.~Norta, ``Design of a remote emotional requirement elicitation
  feedback method,'' in \emph{2020 IEEE Third International Workshop on
  Affective Computing in Requirements Engineering (AffectRE)}.\hskip 1em plus
  0.5em minus 0.4em\relax IEEE, 2020, pp. 3--8.

\bibitem{P23}
T.~Nakatani and K.~Sato, ``Mulsa: Multi-layered scenario analysis for an
  advanced driver assistance system.'' in \emph{ICSOFT}, 2013, pp. 83--91.

\bibitem{P24}
E.~Alatawi, A.~Mendoza, and T.~Miller, ``Psychologically-driven requirements
  engineering: a case study in depression care,'' in \emph{2018 25th
  Australasian Software Engineering Conference (ASWEC)}.\hskip 1em plus 0.5em
  minus 0.4em\relax IEEE, 2018, pp. 41--50.

\bibitem{P25}
S.~A. Scherr, P.~Mennig, C.~Kammler, and F.~Elberzhager, ``On the road to
  enriching the app improvement process with emotions,'' in \emph{2019 IEEE
  27th International Requirements Engineering Conference Workshops
  (REW)}.\hskip 1em plus 0.5em minus 0.4em\relax IEEE, 2019, pp. 84--91.

\bibitem{P26}
Z.~S.~H. Abad, V.~Gervasi, D.~Zowghi, and K.~Barker, ``Elica: An automated tool
  for dynamic extraction of requirements relevant information,'' in \emph{2018
  5th International Workshop on Artificial Intelligence for Requirements
  Engineering (AIRE)}.\hskip 1em plus 0.5em minus 0.4em\relax IEEE, 2018, pp.
  8--14.

\bibitem{P35}
S.~F. Zulkifli, C.~W. Shiang, M.~A. bin Khairuddin, and N.~bt~Jali, ``Modeling
  emotion oriented approach through agent-oriented approach,''
  \emph{International Journal on Advanced Science, Engineering and Information
  Technology}, vol.~10, no.~2, pp. 647--653, 2020.

\bibitem{P28}
J.~Grundy, ``Human-centric software engineering for next generation cloud-and
  edge-based smart living applications,'' in \emph{2020 20th IEEE/ACM
  International Symposium on Cluster, Cloud and Internet Computing
  (CCGRID)}.\hskip 1em plus 0.5em minus 0.4em\relax IEEE, 2020, pp. 1--10.

\bibitem{P29}
E.~Guzman and W.~Maalej, ``How do users like this feature? a fine grained
  sentiment analysis of app reviews,'' in \emph{2014 IEEE 22nd international
  requirements engineering conference (RE)}.\hskip 1em plus 0.5em minus
  0.4em\relax Ieee, 2014, pp. 153--162.

\bibitem{P31}
M.~Levy, ``Emotional requirements for well-being applications: The customer
  journey,'' in \emph{2020 IEEE First International Workshop on Requirements
  Engineering for Well-Being, Aging, and Health (REWBAH)}.\hskip 1em plus 0.5em
  minus 0.4em\relax IEEE, 2020, pp. 35--40.

\bibitem{P32}
D.~Callele, E.~Neufeld, and K.~Schneider, ``Visualizing emotional
  requirements,'' in \emph{2009 Fourth International Workshop on Requirements
  Engineering Visualization}.\hskip 1em plus 0.5em minus 0.4em\relax IEEE,
  2009, pp. 1--10.

\bibitem{P33}
N.~N.~B. Abdullah, J.~Grundy, J.~McIntosh, Y.~C. How, S.~Saharuddin, K.~K. Tat,
  E.~ShinYe, A.~J.~A. Rastom, and N.~L. Othman, ``Using work system design,
  user stories and emotional goal modeling for an mhealth system,'' in
  \emph{2020 IEEE First International Workshop on Requirements Engineering for
  Well-Being, Aging, and Health (REWBAH)}.\hskip 1em plus 0.5em minus
  0.4em\relax IEEE, 2020, pp. 1--10.

\bibitem{P34}
A.~Gregoriades, M.~Pampaka, and M.~Georgiades, ``A holistic approach to
  requirements elicitation for mobile tourist recommendation systems,'' in
  \emph{Future of Information and Communication Conference}.\hskip 1em plus
  0.5em minus 0.4em\relax Springer, 2019, pp. 857--873.

\end{thebibliography}
